\documentclass[preprint2]{aastex}

\usepackage[usenames]{color}

\begin{document}

\title{Automated classification of variable stars in the asteroseismology
program of the \textit{Kepler space mission}}

\author{J. Blomme \altaffilmark{1}, J. Debosscher\altaffilmark{1},
J. De Ridder\altaffilmark{1}, C. Aerts\altaffilmark{1,2}, R.L.
Gilliland\altaffilmark{3}, J. Christensen-Dalsgaard\altaffilmark{4},
H. Kjeldsen\altaffilmark{4}, T.M. Brown\altaffilmark{5}, W.J.
Borucki\altaffilmark{6}, D. Koch\altaffilmark{6}, J.M.
Jenkins\altaffilmark{7}, D.W. Kurtz\altaffilmark{8}, D.
Stello\altaffilmark{9}, I.R. Stevens\altaffilmark{10}, M.D.
Suran\altaffilmark{11}, A. Derekas\altaffilmark{9,12}}

\altaffiltext{1}{Instituut voor Sterrenkunde, Katholieke
Universiteit Leuven, Celestijnenlaan 200D, B-3001 Leuven, Belgium}
\altaffiltext{2}{IMAPP, Department of Astrophysics, Radbout
University Nijmegen, PO Box 9010, 6500 GL Nijmegen, the Netherlands}
\altaffiltext{3}{Space Telescope Science Institute, 3700 San Martin
Drive, Baltimore, MD 21218, USA} \altaffiltext{4}{Department of
Physics and Astronomy, Aarhus University, DK-8000 Aarhus C, Denmark}
\altaffiltext{5}{Las Cumbres Observatory Global Telescope,  Goleta,
CA 93117, USA} \altaffiltext{6}{NASA Ames Research Center, MS
244-30, Moffett Field, CA 94035, USA} \altaffiltext{7}{SETI
Institute/NASA Ames Research Center, MS 244-30, Moffett Field, CA
94035, USA} \altaffiltext{8}{Jeremiah Horrocks Institute of
Astrophysics, University of Central Lancashire, PR1 2HE, UK}
\altaffiltext{9}{Sydney Institute for Astronomy (SIfA), School of
Physics, University of Sydney, NSW 2006, Australia}
\altaffiltext{10}{School of Physics and Astronomy, University of
Birmingham, Edgbaston B15 2TT, United Kingdom}
\altaffiltext{11}{Astronomical Institute of the Romanian Academy,
Str. Cutitul de Argint 5, RO 40557, Bucharest, RO}
\altaffiltext{12}{Konkoly Observatory, Hungarian Academy of Sciences, H-1525 Budapest, P.O. Box 67, Hungary}

%\recieved{}

%\accepted{}

\begin{abstract}
We present the first results of the application of supervised
classification methods to the \textit{Kepler} Q1 long-cadence light
curves of a subsample of 2288 stars measured in the asteroseismology
program of the mission. The methods, originally developed in the
framework of the CoRoT and Gaia space missions, are capable of
identifying the most common types of stellar variability in a
reliable way. Many new variables have been discovered, among which a
large fraction are eclipsing/ellipsoidal binaries unknown prior to
launch. A comparison is made between our classification from
the \textit{Kepler} data and the pre-launch class based on data from
the ground, showing that the latter needs significant improvement.
The noise properties of the \textit{Kepler} data are compared to those
of the exoplanet program of the CoRoT satellite. We find that
\textit{Kepler} improves on CoRoT by a factor 2 to 2.3 in
point-to-point scatter.
\end{abstract}


\section{Introduction}

The \textit{Kepler} satellite, launched in March 2009, is NASA's
first mission capable of finding Earth-size and smaller planets
\citep{Borucki:2010, Koch:2010}. It has an 0.95-m aperture Schmidt
telescope with a photometer comprised of 42 CCDs having a fixed
field of view of 105 square degrees in the constellations Cygnus and
Lyrae. It is designed to monitor continuously the brightness of
$160\,000$ stars during the first year, reduced to $100\,000$ stars
later in the mission. This results in high-quality light curves, not
only interesting for the detection of planets, but also of great
importance for asteroseismology.

In this Letter, we present data from the asteroseismology program of
the \textit{NASA Kepler Mission} \citep{Gilliland:2010}. We present a search
for variable stars and the application of supervised classification
methods to the \textit{Kepler} long-cadence Q1 data of its
asteroseismology program, covering 33.5 days in
total. All the light curves have 29.4-min time sampling. Both the
total time span and sampling are very well suited to study short
period eclipsing and ellipsoidal binaries, classical pulsators such
as RR\,Lyr stars and Cepheids, and nonradial pulsators such as
$\beta\,$Cep stars, Slowly pulsating B stars (SPBs), $\delta$\,Sct
stars and $\gamma\,$Dor stars (see \citet{Aerts:2010} for a definition of
all these classes). We used the stellar fluxes as they
were delivered to us after preliminary data processing
\citep{Jenkins:2010}.
%(KASOC\footnote{http://kasoc.phys.au.dk/kasoc/}).
In total, we analyzed 2288 \textit{Kepler} light curves.

We compare our results to a pre-launch classification based on data in
the \textit{Kepler Input Catalog} (\textit{KIC} hereafter) and prepared
by the \textit{Kepler Asteroseismic Science Consortium} (KASC, Gilliland et al. 2010).
Finally, we compare the point-to-point scatter of the \textit{Kepler} data to
the one of CoRoT's exoplanet data of the first long run (5 months)
of that mission.

\section{Adopted Methodology}

To detect and extract the variables, we relied on the automated
variability characterization method as described in detail in
\citet{PaperI,PaperIII}. This method searches for three independent
frequencies for every star, which are used to make a harmonic
best-fit to the trend-subtracted time series. In this way we obtain
a homogeneous set of parameters for each star, irrespective of its
variability nature. The goal is \textit{not} to achieve a good light
curve model, but rather to deduce a set of light curve parameters
which is sufficient to classify the variability.

Using this set of
parameters, we classified the stars using a modified version of the
classifier based on Gaussian mixtures, described in \citet{PaperI},
with the definition stars from \citet{PaperIII}. We improved the
performance of the algorithm presented in \citet{PaperIII}
by classifying the objects using a multi-stage tree. In
each node, we decide which groups of stars we want to distinguish.
The best parameters are then selected for that node and for each
group the parameter distribution is approximated with a mixture of
multi-dimensional Gaussians. To each variable stellar target we
assign a probability that it belongs to a particular group.

In order to obtain the final probability for each variability class we multiply
the probabilities along the corresponding root-to-leaf path (see, e.g.,
\citet{Ripley:1996} for a general introduction and definition of tree-structured
classifiers). In practice, this works as follows for our application: the
probability of a stellar target being an RR\,Lyr star of type ab is, e.g., the
probability of not being an eclipsing binary (first stage) times the probability
of belonging to the RR\,Lyr group (second stage) times the probability of being
an RR\,Lyr star of type ab.

This procedure was followed to compute class probabilities for each
target. In order to obtain the best candidates, we additionally used
the Mahalanobis distance, which is a multi-dimensional
generalization of the one-dimensional statistical or standard
distance as described in \citet{PaperIII}. A visual check has been
performed as well.

\section{Classification results}

The variability classes we currently take into account, and the
number of good candidate class members, are listed in
Table\,\ref{classes}. While we classified more than 200 stars
securely, the majority of targets still has a too ambiguous class
assignment, mainly due to the limited time base and to the large
fraction of red giants among the sample (see below). Nevertheless,
we managed to identify numerous new pulsators and binaries from the
short time series and from early data reduction.  All the
illustrations presented in this paper contain stars that were either
unknown as variables or were misclassified prior to launch.

%The results of our classification have been implemented in the KASOC database.

\begin{figure*}
\includegraphics[width = 16.0cm]{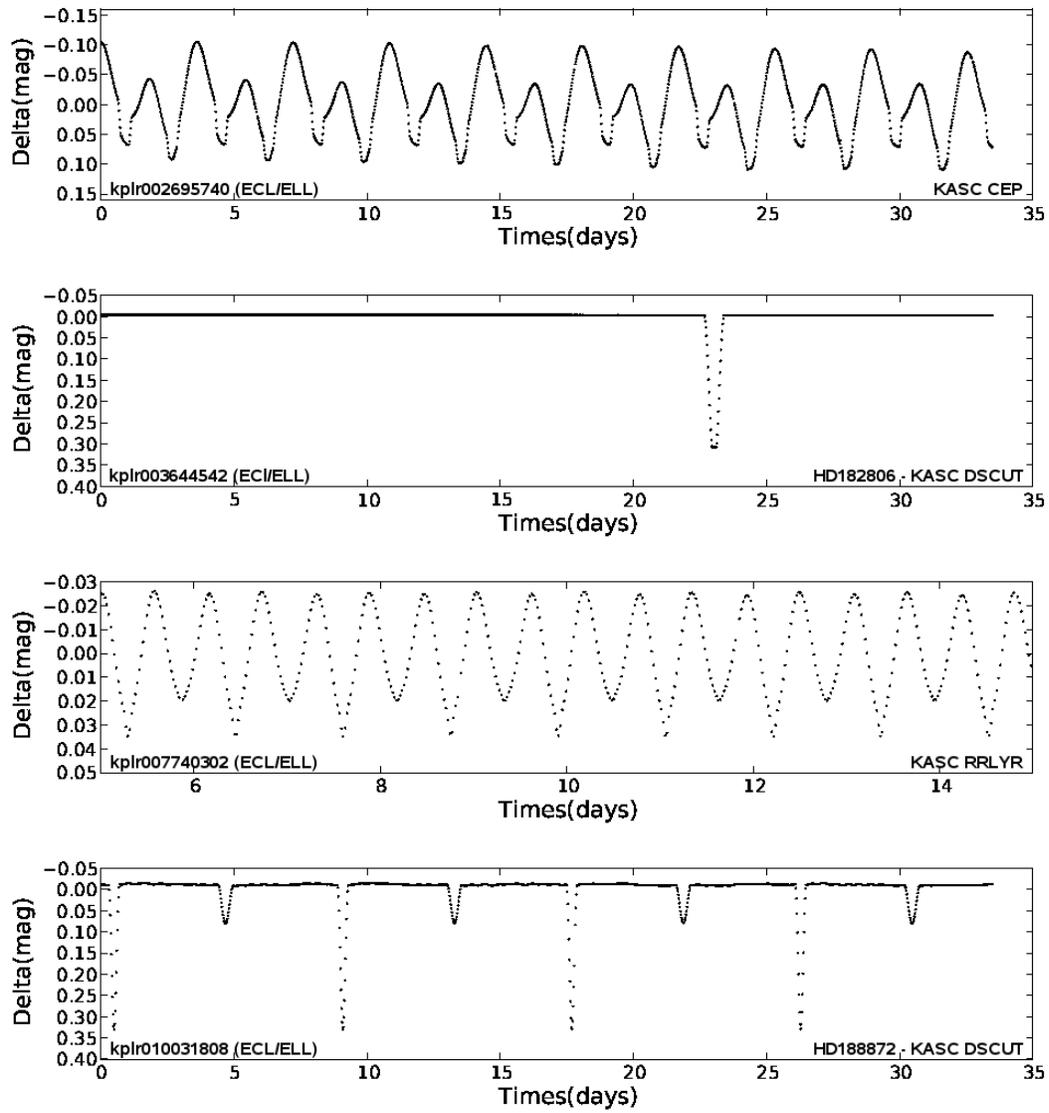}\\
\caption{Some newly discovered eclipsing binaries, which were not
identified as such prior to launch. The pre-launch KASC class
is given on the right.} \label{lc_ecl}
\end{figure*}

We evaluated our classification results by visual inspection and
manual analysis of the best candidates based on the Mahalanobis
distance, and by comparing them with the pre-launch classification
which is often insecure due to limited ground data. The results are
summarized in Table\,\ref{classes}. In the second and third column,
a Mahalanobis distance smaller than 3 and 2 is taken, respectively,
without visual inspection. In the fourth column the final numbers of
candidates are given, taking a Mahalanobis distance less than 3 and
performing an additional visual inspection. In the last column, the
pre-launch classification done by KASC members is given. Note that
several targets occur in more than one pre-launch class. For the
majority of the classes in Table\,\ref{classes}, our results
appreciably improve the pre-launch results, which were necessarily
based on ground-based data and could not take into account
\textit{Kepler}'s high quality light curves. Good agreement is
obtained for binaries and classical pulsators such as RR\,Lyr stars
and Cepheids, but even there, we are able to improve the pre-launch
results as some RR Lyr candidates were reclassified by us as
eclipsing binaries. One example is given in the third panel of
Fig.\,\ref{lc_ecl}.

We could also identify many new (eclipsing) binaries, some of which are shown in
Fig.\,\ref{lc_ecl}.  An example of a red giant pulsator with solar-like
oscillations in an eclipsing binary is discussed in detail in
\citet{Hekker:2010}.

\begin{figure*}
\includegraphics[width = 16.0cm]{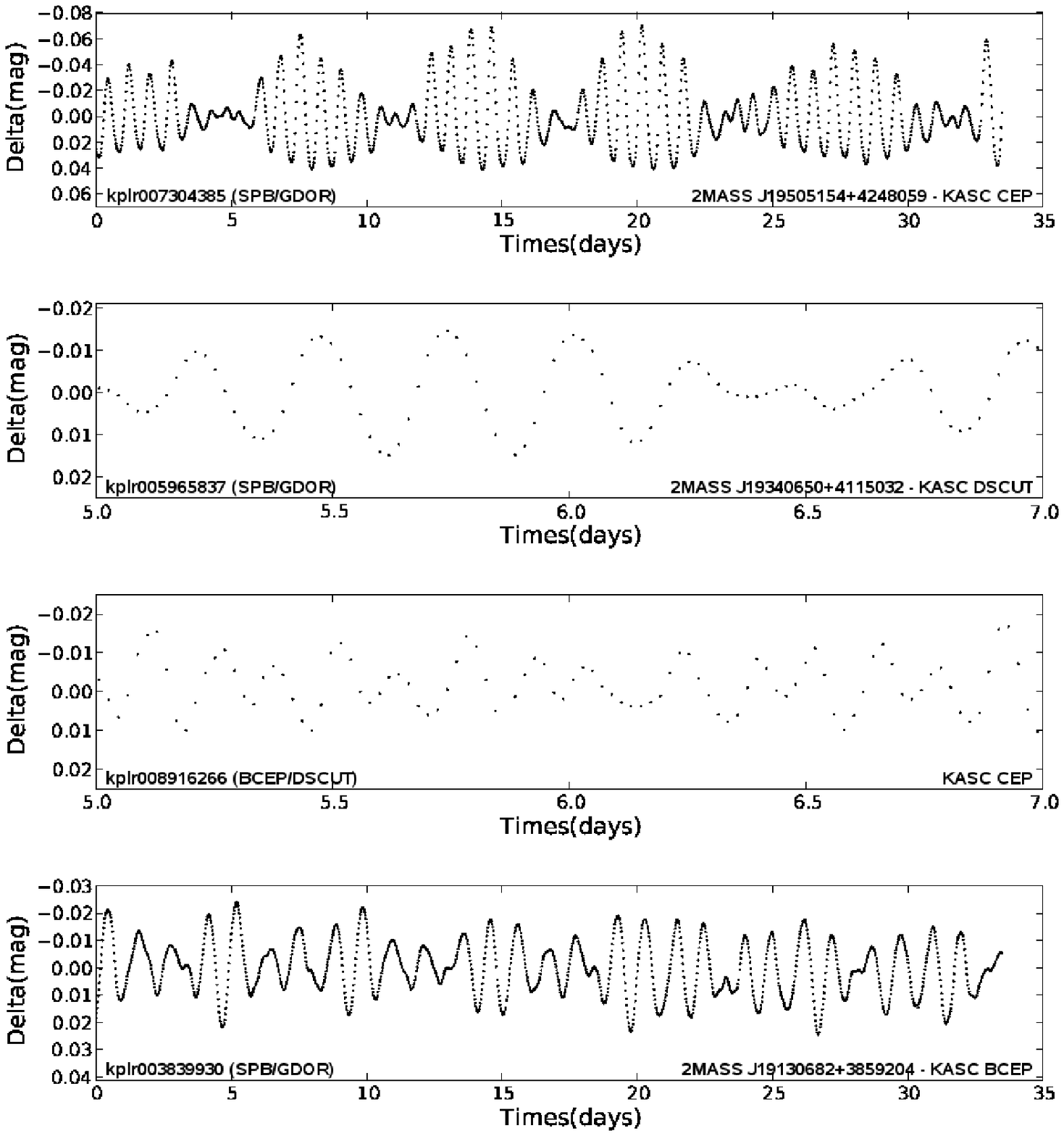}\\
\caption{Some examples of multiperiodic pulsators, not present in the pre-launch
class lists of pulsators. The pre-launch KASC class is given on the right.}
\label{lc_multivar}
\end{figure*}

For main-sequence nonradial pulsators, such as $\beta$ Cep, $\delta$
Sct, SPB and $\gamma$ Dor stars, there is much discrepancy between
the pre-launch and our classification. Few of the pre-launch
candidates turn out to be actual class members. Indeed, we do not
find high-probability candidates among these stars with a pre-launch
class assignment.  On the other hand, we identified new nonradial
pulsators not present in the pre-launch lists. Some examples are
shown in Fig.\,\ref{lc_multivar}. Similar light curves, albeit for
fainter stars, were found in the CoRoT exoplanet database
\citep[e.g.]{Degroote:2009}.

We have not yet been able to compare our results for the long period variables
along or past the Asymptotic Giant Branch, such as Miras or RV\,Tau stars, since
the current total time span of the light curves is only a fraction of the
typical pulsation periods of those objects. Moreover, we did not yet search for
short-period pulsators, such as solar-like pulsators along the main sequence,
rapidly oscillating Ap stars, subdwarf OB variables and white dwarf pulsators
(see \citet{Aerts:2010} for class definitions), given that we do not yet have
short-cadence data.

Another point of attention is the classification of solar-like
pulsators. Stochastic pulsations are more easily recognized from a
broad power excess than from the methodology adopted here. Indeed,
the automated selection of the three highest frequency peaks will
almost always be peaks due to granulation and/or background noise.
Thus, to find such pulsators, one better uses an extractor-type
approach.  This involves fitting and subtraction of the granulation
and background signal to characterize the type of star, after which
the oscillations can be sought.  We refer to \citet{Chaplin:2010}
, \citet{Bedding:2010} and \citet{Stello:2010} for a study of solar-like pulsators among
\textit{Kepler} targets.

Finally, we stress that our classifiers solely use the information
contained in the \textit{Kepler} light curves. This implies that we
cannot discriminate well between the class pairs of B-type $\beta$
Cep and A-type $\delta$ Sct stars with periodicities of the order of
hours, and B-type SPB versus F-type $\gamma$ Dor stars with
oscillation periods of days. Both pairs of classes contain light
curves with very similar characteristics. To discriminate between
them, at least some spectral information, such as a properly
dereddened B-V color index or a stellar spectrum, is needed.
According to the  Initial Mass Function, most of these candidates
should be AF-type pulsators.

\begin{deluxetable}{lllll}
\tablewidth{0pc} \tabletypesize{\footnotesize} \tablecaption{Stellar
variability classes considered in this work. See Chapter 2 in
\citet{Aerts:2010} for a definition of the classes. MD stands for
the Mahalanobis distance as defined in \citet{PaperIII}. A hyphen
`-' indicate variability types not taken into account in our
classification scheme.}
\tablehead{\colhead{Stellar variability classes} &  \colhead{MD $<$ 3} & \colhead{MD $<$ 2} & \colhead{MD $<$ 3}& \colhead{KASC}\\
\colhead{ } &  \colhead{ } & \colhead{ } & \colhead{+ visual inspection}& \colhead{pre-launch class}}
\startdata
 Mira variables (MIRA)                    & 0   &   0 &   0 & 305 \\
 RV-Tauri stars (RVTAU)                   & 1   &   1 &   0 &   2 \\
 Cepheids (CEP)                       & 7   &   3 &   6 &  35 \\
 Compact pulsators (Compact)              & -   &   - &   - &   4 \\
 RR-Lyr stars (RRLYR)                 & 28  &  28 &  28 &  51 \\
 $\beta\,$Cep or $\delta\,$Sct stars (BCEP/DSCUT) & 65  &  15 &  28 &  57 \\
 SPB or $\gamma\,$Dor stars (SPB/GDOR)        & 28  &  13 &  28 &  20 \\
 Ellipsoidal variables (ELL)              & 31  &   8 &  23 &   0 \\
 Eclipsing binaries (ECL)                 & 101 &  92 & 101 & 100 \\
 Red giants (RG)                  & -   &   - &   - & 1513\\
 Rapidly oscillating Ap stars (ROAP)          & -   &   - &   - &   4 \\
 Oscillations in clusters (OS-clus)       & -   &   - &   - & 178 \\
 Solar-like oscillations (SOL)            & -   &   - &   - &  73 \\
\enddata
\label{classes}
\end{deluxetable}

\section{Noise properties of the \textit{Kepler} data in the asteroseismology program}

\begin{figure}
\includegraphics[angle=270,scale=.28]{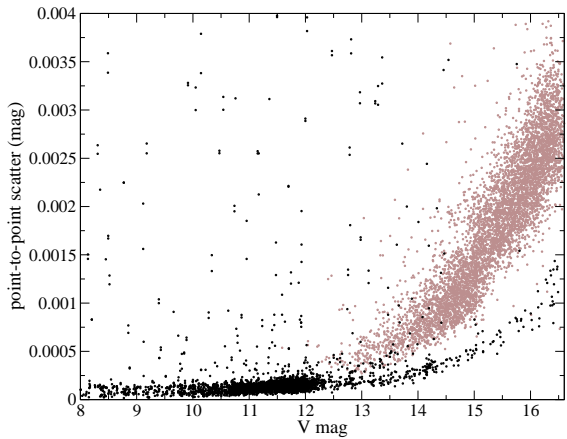}
\caption{Point-to-point scatter of the long-cadence light curves in
\textit{Kepler}'s asteroseismology program \citep[black]{Gilliland:2010} compared
to the one of the light curves in the CoRoT LRc01 exoplanet programme
\citep[gray]{Auvergne:2009}} \label{ptp}
\end{figure}

Figure\,\ref{ptp} reveals the point-to-point scatter in the
\textit{Kepler} data of its asteroseismology program as described in
\citet{Gilliland:2010}, as well as a comparison with that for the
CoRoT space mission's exoplanet data for its first long run of 5
months \citep{Auvergne:2009,Aigrain:2009}. We rescaled the
CoRoT data to the same integration time of 29.4\,min as for the
long-cadence mode of \textit{Kepler} in order to ensure an
appropriate comparison. Duty cycles of these CoRoT and
\textit{Kepler} data are some 90\% and 99\%, respectively. It can be
seen that the \textit{Kepler} data with still preliminary processing
\citep{Jenkins:2010} already outperforms the CoRoT exoplanet data by
a factor $\sim$2 for objects with a visual magnitude of around 14 to
a factor of $\sim$2.3 for objects of magnitude around 16. This is
more or less as expected based on the difference in aperture size of
the two instruments. Importantly for follow-up studies of the most
interesting variable stars, the \textit{Kepler} asteroseismology
sample focuses on brighter objects.

Typical noise levels of the least variable stars in the
\textit{Kepler} asteroseismology program, estimated as the average
amplitude in the Fourier transform avoiding the low-frequency regime
below 1 cycle per day, range from 1.3 $\mu$mag for an 8th magnitude
star to 34 $\mu$mag for a 16th magnitude star. More than 70\,\% of
the KASC stars are variable in one way or another, even taking into
account residual instrumental effects. Note that this is a much
higher fraction than for the CoRoT exoplanet programme, because the
KASC target selection was aimed at focusing on variable stars while
the CoRoT exoplanet sample is unbiased with respect to variability.

\section{Conclusions}

We presented the first results of the application of automated
supervised classification methods to 2288 \textit{Kepler} light
curves.  Comparison with existing pre-launch classification and
manual classification of the light curves shows the capabilities of
our methodology: we are able to improve the pre-existing
classification results seriously, and to identify new class members,
unknown prior to launch.

We will repeat the classifications as more and longer time-span
light curves become available. This way, we will also be able to
identify variables with longer periodicities and have much better
capacity to unravel beat periods in multiperiodic pulsators. More
classes will be included in the classification scheme, and, even
more importantly, the class definition stars will be updated using
the high quality \textit{Kepler} light curves themselves.  Access to
1-min cadence data will allow us to classify shorter period
variables, in additional to those presented here. With each new set
of data, our results will be updated and made available.

\acknowledgments We gratefully thank the entire \textit{Kepler}
team, whose excellent efforts have made these results possible.
Funding for this Discovery mission is provided by
NASA's Science Mission Directorate. The research leading to these
results has received funding from the European Research Council
under the European Community's Seventh Framework Programme
(FP7/2007--2013)/ERC grant agreement n$^\circ$227224 (PROSPERITY),
as well as from the Research Council of K.U.Leuven grant agreement
GOA/2008/04 and from the Belgian Federal Science Office. DWK
acknowledges support by the UK Science and Technology Facilities
Council.

{\it Facilities:} \facility{The Kepler Mission}

%\bibliographystyle{apj}
%\bibliography{references-classification}

\begin{thebibliography}{}
\bibitem[Aerts et al.(2010)]{Aerts:2010} Aerts C., Christensen-Dalsgaard, J., \& Kurtz, D.W. 2010, Asteroseismology, Springer-Verlag (ISBN 978-1-4020-5178-4)
\bibitem[Aigrain et al.(2009)]{Aigrain:2009} Aigrain, S., et al. 2009, \aap, 425, 429
\bibitem[Auvergne et al.(2009)]{Auvergne:2009} Auvergne, M., et al. 2009, \aap, 506, 411
\bibitem[Bedding et al.(2010)]{Bedding:2010} Bedding, T.R., et al. 2010, \apj, submitted
\bibitem[Borucki et al.(2010)]{Borucki:2010} Borucki, W.J., et al. 2010, PASP, in press
\bibitem[Chaplin et al.(2010)]{Chaplin:2010} Chaplin, B.W., et al. 2010, \apj, submitted
\bibitem[Debosscher et al.(2007)]{PaperI} Debosscher, J., Sarro, L.M., Aerts, C., Cuypers, J., Vandenbussche, B., Garrido, R., \& Solano, E. 2007, \aap, 475, 1159
\bibitem[Debosscher et al.(2009)]{PaperIII} Debosscher, J., et al. 2009, \aap, 506, 519
\bibitem[Degroote et al.(2009)]{Degroote:2009} Degroote, P., et al. 2009, \aap, 506, 471
\bibitem[Gilliland et al.(2010)]{Gilliland:2010} Gilliland, R.L., et al. 2010, PASP, in press
\bibitem[Hekker et al.(2010)]{Hekker:2010} Hekker, S., et al. 2010, \apj, submitted
\bibitem[Jenkins et al.(2010)]{Jenkins:2010} Jenkins, J.M., et al. 2010, \apj, submitted
\bibitem[Koch et al.(2010)]{Koch:2010} Koch, D., et al. 2010, PASP, in press
\bibitem[Ripley et al.(1996)]{Ripley:1996} Ripley, B.D. 1996, Pattern Recognition and Neural Networks, Cambridge University Press
\bibitem[Stello et al.(2010)]{Stello:2010} Stello, D., et al. 2010, \apj, submitted
\end{thebibliography}

\end{document}